\documentclass[final,5p,authoryear,times]{elsarticle} 
\pdfoutput=1

\usepackage{graphicx}
\usepackage{amsmath}
\usepackage{booktabs}
\usepackage{colortbl} 
\usepackage{amssymb}
\usepackage{makecell}
\usepackage{color}

\usepackage[T1]{fontenc}
\usepackage{ae,aecompl}
\usepackage{enumitem}
\usepackage{times}
\usepackage[hidelinks]{hyperref}
\usepackage{xcolor}
\usepackage{subfig}
\usepackage[ruled,vlined]{algorithm2e}
\usepackage[utf8]{inputenc}
\usepackage{tabularx}
\usepackage{amsmath}
\usepackage{adjustbox}

\DeclareMathOperator*{\argmin}{arg\,min}

\hypersetup{
    colorlinks,
    linkcolor={red!50!black},
    citecolor={blue!50!black},
    urlcolor={blue!80!black}
}
\usepackage{multirow}
\usepackage{mwe}
\usepackage{rotating}
\usepackage[normalem]{ulem}
\useunder{\uline}{\ul}{}
\usepackage{dblfloatfix}
\usepackage{microtype}

\newcommand{\rthis}[1]{\textcolor{black}{#1}}
\DeclareUnicodeCharacter{2212}{-}
\journal{Astronomy \& Computing}
\begin{document} \sloppy
\begin{frontmatter}

\title{Galaxy Morphology Classification using Neural Ordinary Differential Equations}

\author[1]{Raghav Gupta}\ead{cs19mtech11024@iith.ac.in}
\author[1]{P.K. Srijith}\ead{srijith@iith.ac.in}
\author[2]{Shantanu  Desai}\ead{shantanud@phy.iith.ac.in}
\address[1]{Department of Computer Science and Engineering, IIT Hyderabad, Kandi,  Telangana-502285, India}
\address[2]{Department of Physics, IIT Hyderabad, Kandi,  Telangana-502285, India}




\begin{abstract} \label{abstract}
We introduce a continuous depth version of the Residual Network (ResNet) called  Neural ordinary differential equations (NODE) for the purpose of galaxy morphology classification. We carry out a classification of  galaxy images from the Galaxy Zoo 2 dataset, consisting of  five distinct classes, and obtained an accuracy between 91-95\%, depending on the image class. We train NODE with different numerical techniques such as  adjoint and Adaptive Checkpoint Adjoint (ACA) and compare them against ResNet.
While ResNet has certain drawbacks, such as time consuming architecture selection (e.g. the number of layers) and the requirement of a large dataset needed for training, NODE can  overcome these limitations.  Through our results, we show that that  the accuracy of NODE is comparable to  ResNet, and the  number of parameters used is about one-third as compared to ResNet, thus leading to a smaller memory footprint, which would benefit  next generation  surveys.


\end{abstract}

\end{frontmatter}

\section{INTRODUCTION} 
\label{INTRODUCTION}
\par 

The problem of determining the  morphology of a galaxy plays a pivotal role in a large number of fields from galaxy evolution to cosmology.  Some of these applications include  stellar masses~\citep{Bundy}, star formation history~\citep{Kennicutt}, color~\citep{Skibba},  gas and dust content~\citep{Lianou}, age of the galaxy~\citep{Bernardi}, various dynamical processes~\citep{Fall}, tests of modified gravity theories~\citep{Pedro} etc. A recent review on  various aspects of  galaxy morphology and its connections to the rest of astrophysics can be found in ~\citet{Buta}.

The very first morphological classification schemes pioneered by \citet{Hubble26}  were based upon  visual scanning of  galaxies and classifying them into different types such as spirals, ellipticals, lenticulars. With the advent of large area optical surveys, the task of visual classification was outsourced to the Galaxy Zoo project~\citep{Lintott}. The first incarnation of the project (Galaxy Zoo 1), consisting of  a dataset of more than 900,000 images by the Sloan Digital Sky Survey~\citep{York}, was classified by citizen scientists    into four categories:  “spiral”, “elliptical”, “a merger” or “star/don’t know”~\citep{Lintott}. The project enabled the annotation of a million galaxy images  within several months. This was superseded by Galaxy Zoo 2~\citep{Willett2013}, Galaxy Zoo: Hubble~\citep{Galaxyzoohubble}, and Galaxy Zoo: CANDELS~\citep{Galaxyzoocandels}.

Unfortunately, this manual approach of visual classification  does not scale well   with the unprecedented pace of data growth due to the large number of meter-class telescopes equipped with multi-CCD imagers, which have been continuously  built over the past two decades. Very soon stage IV Dark Energy surveys such as  Legacy Survey of Space and Time  operated by the Vera Rubin observatory~\citep{LSST}, Euclid~\citep{Euclid}, and Roman Space Telescope~\citep{WFIRST}  are going to produce petabytes worth of data, rendering manual classification impossible.



Therefore, astronomers have turned their attention
 to automated classification methods. Over the past few decades, a large amount of literature has emerged on such automated  methods for measuring galaxy morphology, especially in large observational surveys.  These methods range from parametric techniques, which attempt to describe the  galaxy light profiles using small sets of parameters~\citep{simard2002deep,Sersic1963InfluenceOT,odewahn2001automated,lackner2012astrophysically}, to non-parametric methods that reduce these light distributions to single values such as in the ‘CAS’ system~\citep{conselice2003relationship,Abraham,menanteau2005morphological}, the Gini-M20 coefficients~\citep{lotz2003new,freeman2013new}, etc. Recent reviews of some of these automated methods can be found in \citet{Diego,Kaviraj}.
 
 A major game changer throughout astronomy and astrophysics has been the widespread application of machine learning and deep learning techniques~\citep{Brunner09,Kremer,Bethapudi,Baron19}, and galaxy morphology is no exception to this. Applications of machine learning as well as deep learning  to galaxy morphology classifications are discussed  in ~\citet{Dieleman2015,Drlica,Tuccillo,Barchi,Khan,Spindler,Bhambra21,Reza}.

Deep learning models, known as deep neural networks (DNN),  have been widely used for image classification and slowly began to beat human accuracy in these tasks, as soon as  large training sets started becoming available~\citep{dnn}. 
DNN models, especially Convolution Neural Networks (CNN) \citep{cnn},  AlexNet, VGGNet and GoogleNet, took the accuracy of DNNs to new heights. With the advent of Residual Networks (ResNet) \citep{DBLP:journals/corr/HeZRS15}, researchers were able to make these CNN models deeper than ever before, without suffering from additional problems.
Among the machine learning techniques,  CNNs~\citep{cnn} have become the mainstream method for image classification. However, CNN with a large number of layers  suffer from the  vanishing gradient problem~\citep{vangrad}.

In the popular deep learning models such as ResNets, the selection of architecture (depth of the network) and the presence of a large number of parameters can make the training process computationally intractable. 
Recently, a continuous depth counterpart to ResNets,  known as NODE~\citep{NODE} was introduced, which could overcome these drawbacks.  In our work, we propose to use NODE for the galaxy morphology classification problem. We  compare its performance against ResNet, which has also been used  in other works~\citep{dai2018galaxy,Goddard}, as that  is  the state-of-the-art deep learning approach for galaxy morphology classification and demonstrate the  benefits of NODE over ResNets. 

NODE is inspired by the way ResNet works, where one models the change in the feature maps over layers using a neural network. This can be seen as equivalent to an ordinary differential equation with the derivative modelled as a  neural network function.  Consequently, the final layer feature map can be obtained using numerical solvers for ODE such as Euler's method  and Runge-Kutta method. NODE has certain advantages over ResNet. In NODE, the network depth is implicitly determined by the tolerance parameter of the numerical solver used, rather than being explicitly fixed like in ResNet. Thus, by tuning the tolerance parameter, we can  trade-off between the model speed and model accuracy. Another advantage of NODE over ResNet is that the number of parameters in NODE is much less than ResNet. Models with smaller number of parameters require less data to train and do not suffer from over-fitting issues. With new training techniques emerging in this field, like Adaptive Checkpoint Adjoint~\citep{NODE_ACA}, NODE architecture is becoming more accurate and faster with time.
  
The NODE architecture has been applied to a  wide variety of fields, such as  biomedical imaging, high-energy physics, image and video processing, 3D modelling, economics, etc~\citep{groha2020neural}. For example, in the case of biomedical-imaging, it has been used for kidney segmentation \citep{valle2019neural}, reconstruction of MRI images \citep{chen2020mri}, multi-state survival analysis \citep{groha2020neural},   3-D modelling for accurate manifold generation \citep{gupta2020neural}, small-footprint keyword spotting (KWS) in audio files \citep{fuketa2020neural}, etc. In the domain of theoretical High-Energy Physics, it has been  applied to holographic QCD~\citep{hashimoto2020neural}. However, to the best of our knowledge, this technique has not been previously applied to any problem in astrophysics.


The organization of this manuscript is as follows. In Section~\ref{DATASET}, we describe the dataset used to carry out our experiments. Next, we shed some light on ResNet  in Section~\ref{RESNET}, followed by an in-depth explanation of the working of NODE (Section~\ref{NODE}) and its training with the adjoint method. 
 We describe the various pre-processing steps  applied to the data, followed by the  exact network architecture used in Section~\ref{EXPERIMENTAL SETUP}. Then, in Section \ref{RESULTS AND DISCUSSION}, we discuss our experimental results. Finally, we conclude in Section \ref{CONCLUSIONS}.


\section{DATASET} 
\label{DATASET}
\begin{figure}
\centering
\label{fig:classes_images}
\begin{tabular}{ccccc}
\subfloat{\includegraphics[width = 0.8in, height=0.5in]{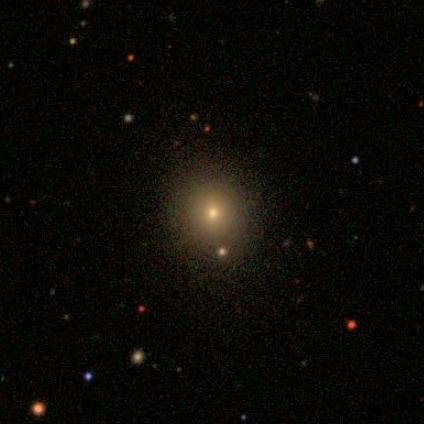}} &
\subfloat{\includegraphics[width = 0.8in, height=0.5in]{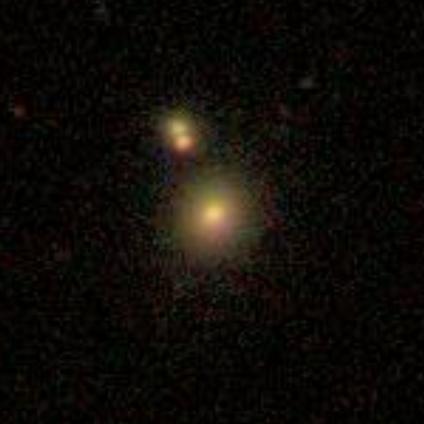}} &
\subfloat{\includegraphics[width = 0.8in, height=0.5in]{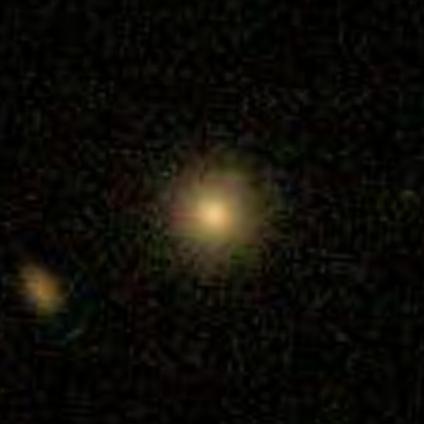}} \\
\subfloat{\includegraphics[width = 0.8in, height=0.5in]{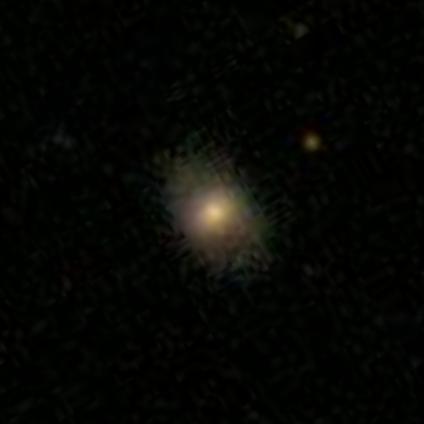}} &
\subfloat{\includegraphics[width = 0.8in, height=0.5in]{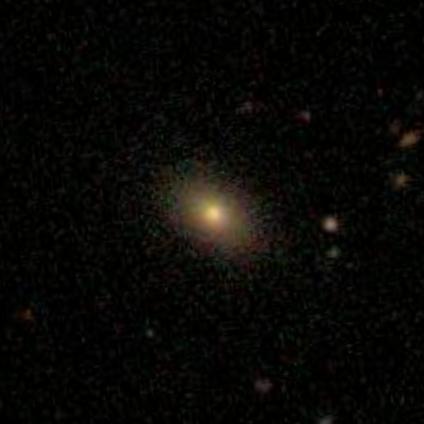}} &
\subfloat{\includegraphics[width = 0.8in, height=0.5in]{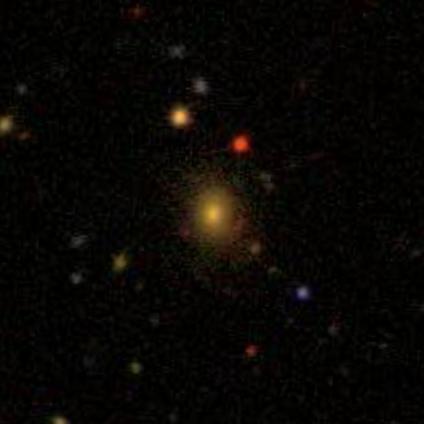}}\\
\subfloat{\includegraphics[width = 0.8in, height=0.5in]{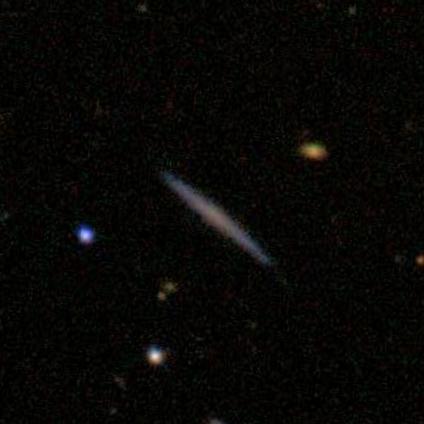}} &
\subfloat{\includegraphics[width = 0.8in, height=0.5in]{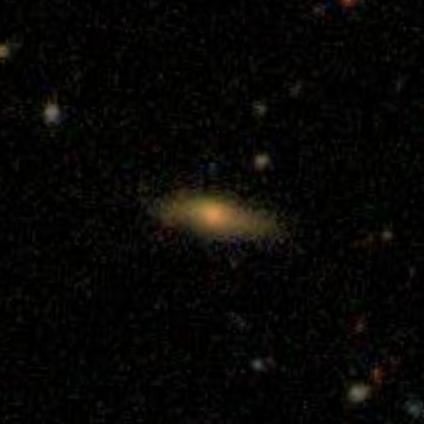}} &
\subfloat{\includegraphics[width = 0.8in, height=0.5in]{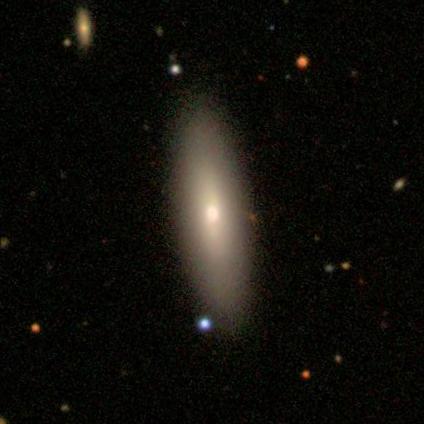}}\\
\subfloat{\includegraphics[width = 0.8in, height=0.5in]{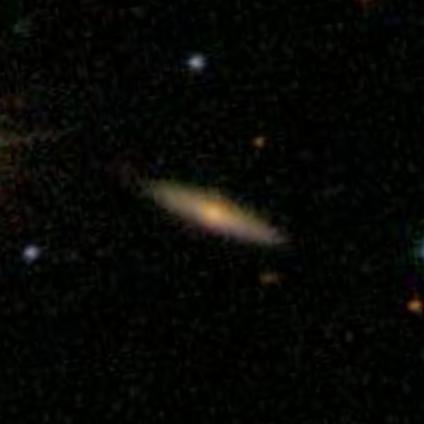}} &
\subfloat{\includegraphics[width = 0.8in, height=0.5in]{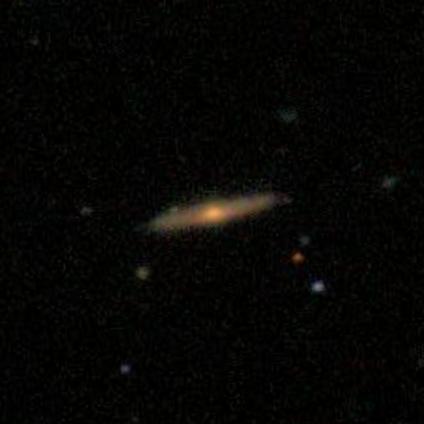}} &
\subfloat{\includegraphics[width = 0.8in, height=0.5in]{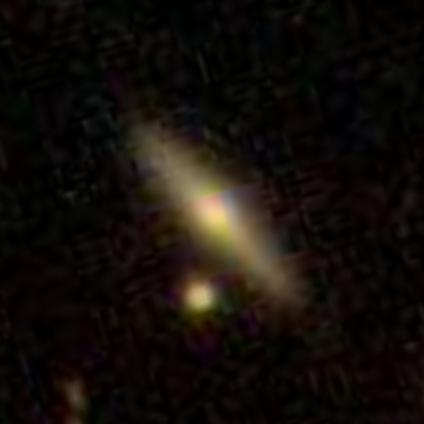}}\\
\subfloat{\includegraphics[width = 0.8in, height=0.5in]{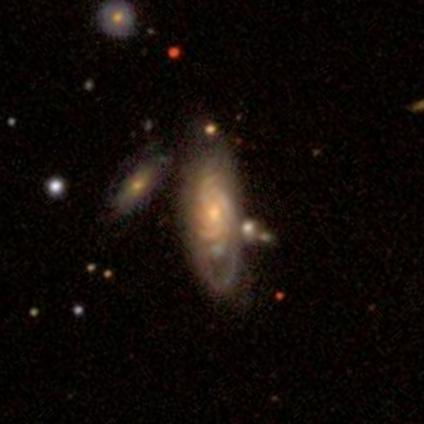}} &
\subfloat{\includegraphics[width = 0.8in, height=0.5in]{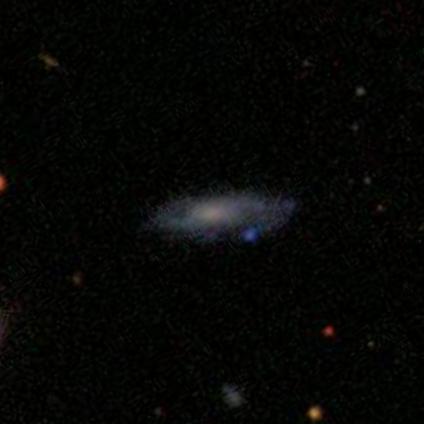}} &
\subfloat{\includegraphics[width = 0.8in, height=0.5in]{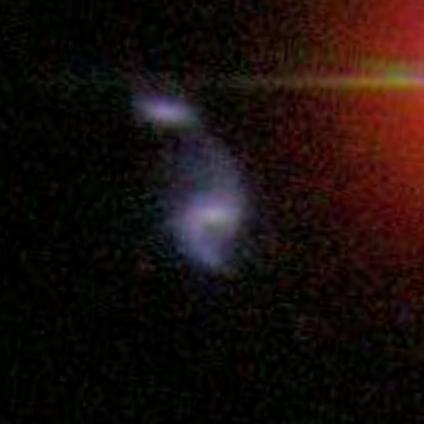}}
\end{tabular}
\caption{The five different galaxy morphologies in the Galaxy Zoo-2 dataset. These classes are completely round smooth, in-between smooth, cigar-shaped
smooth, edge-on and spiral, from top to bottom. See also Fig.~1 of ~\citet{dai2018galaxy} for more examples of different galaxy morphologies from this dataset.}
\label{fig:classes}
\end{figure}

 The dataset used in our experiments is drawn from the Galaxy Zoo Challenge, available on kaggle. Classification labels for the kaggle Dataset (KD) are drawn from Galaxy Zoo 2, and the  images used were obtained from SDSS-DR7~\citep{DR7}. Galaxy images used in this dataset are classified into a  total of five classes viz. spiral, edge-on, cigar-shaped smooth, in-between smooth, and completely round smooth. The different morphological types are shown in Fig.~\ref{fig:classes}. Similar to ~\citet{dai2018galaxy}, we shall use the numerical labels 0, 1, 2, 3, 4 to annotate completely round, in-between, cigar-shaped, edge-on, and spiral galaxies, respectively.

KD consists of around 60,000 images, and  each image is divided into five classes, with a classification probability provided for each class. We prune this dataset further and only select those images, which are classified with high probability in their respective classes. After pruning, we are left with a total of 28,790 images, with a single class assigned to each image. \rthis{We should however point out that there is no absolute ground truth but rather only the truth as estimated via
crowdsourcing.}

This selection criteria is similar to that described in \citet{Willett2013}, in which the galaxy images classified with  probabilities higher than a certain threshold (discussed therein), are selected.
After these cuts, we have 7806, 3903, 578, 8069, and 8434 images in each class, in the order listed at the beginning of this section.

The size of each image is  $424\times424\times 3$ pixels, where the last dimension denotes the number of color channels viz. RGB. The galaxy of interest is generally located at the center of the image. We finally split  our data randomly in the ratio of 9:1 for the purpose of training and testing, thus assigning 25911 and 2879 galaxies, respectively for each task similar to \citet{dai2018galaxy}. 
We create multiple random train and test splits and obtain the average and variance across them in order to conduct a more robust evaluation and  to obtain error estimates on our machine learning metrics.

\section{Residual Neural Networks} 
\label{RESNET}


Neural networks are modelled as a series of transformations having discrete number of layers, with  each one taking in a previous hidden representation  $\mathbf{h}_l$ and producing a new hidden representation $\mathbf{h}_{l+1} = F^l(\mathbf{h}_{l})$. 
We typically consider the transformation as $F(\mathbf{x}) = \sigma(\sum_{i}W_i x_i)$, where $\sigma$ is an activation function (e.g. RELU or a sigmoid), and $\theta$ is a collection of  weight vectors. Recently, many deep learning models were introduced based on the idea that increasing the number of layers can improve the performance. However, this may lead to problems such as vanishing gradients~\citep{vangrad}, where the initial layer weight vectors cannot be computed correctly through back propagation as the error gradient becomes small.

This problem was addressed by Deep Residual Learning~\citep{DBLP:journals/corr/HeZRS15}. ResNets are  a class of  DNNs, which try to map the residuals instead of the complete transformation itself in the hidden layer mappings.  The idea  is to learn a mapping as the difference between the layers (or equivalently adding skip connections): $\mathbf{h}_{l+1} = F^l(\mathbf{h}_l) + \mathbf{h}_l$. In \citet{DBLP:journals/corr/HeZRS15}, they showed that this simple transformation avoids the  vanishing gradient problem due to skip connections and the networks can learn the weights properly.  This allowed the development of deep learning models with a large number of layers (e.g. ResNet with 50 and 100 layers)

ResNet and its variants were able to achieve state of the art results for image classification. ResNet  won the ILSVRC challenge in 2015. Many other variants of ResNet, achieved state-of-the-art (SOTA) results in other image datasets. ResNet mainly has two types of residual blocks.
In the standard block, two 3$\times$3 convolutions are applied, along with a skip connection. In another block, known as the bottleneck block, 1$\times$1 convolutions are applied before and after the 3$\times$3 convolutions, in order to reduce feature space, so that the computational complexity is reduced.

\section{Neural Ordinary Differential Equations} 
\label{NODE}

\subsection{Residual networks}
Recently,~\citet{NODE, lu2017finite} have shown that continuous depth ResNets, known as NODE,  can be developed by relating them to ordinary differential equations. Assuming the mapping function to be the  same across all the layers, and letting $\Delta t \in \mathbb{R}$,  we can rewrite the hidden representation update of ResNets as state updates at some time $t$.
\begin{equation} \mathbf{h}(t+1) = F(\mathbf{h}(t)) + \mathbf{h}(t) = \frac{\Delta t}{\Delta t} F(\mathbf{h}(t)) + \mathbf{h}(t) \\ \\= \Delta t
G(\mathbf{h}(t)) + \mathbf{h}(t) 
\label{eqn:1}
\end{equation}
where $G(\mathbf{h}(t)) = F(\mathbf{h}(t))/\Delta t$. This reformulation is the  same as the single step of Euler's method for solving ordinary differential equations of the form as observed in \citet{lu2017finite}. 
\begin{equation}
\frac{d \mathbf{h}(t)}{dt} = G(\mathbf{h}(t), t, \theta)
\label{eqn:node}
\end{equation}

As compared to standard differential equations, the derivative is represented by a function parameterized using a neural network $G$ acting on the state $\mathbf{h}(t)$. Here, we have assumed the $G$'s to depend on $t$ as well as some parameters $\theta$ (parameters of the neural network). One can consider $G$  to represent convolution operation when applied to the image data. Considering Eq.~\eqref{eqn:1}, the final representation (feature map) of our network is the state $\mathbf{h}(T)$ at time $T$. This is then fed to a fully-connected neural network (FCNN) to predict the final output, which is a real number for regression problems and a discrete value for classification.
For a neural network function $G$, we can use any off-the-shelf ODE solvers such as Euler and Runge-Kutta (RK4) method to solve and obtain the final representation in an iterative manner. 
$$\mathbf{h}(T) = \text{ODESolve}(\mathbf{h}(t_0), G, t_0, T, \theta)$$


 

\subsection{Training Process}
Training a NODE involves learning the parameters of the neural network function using an appropriate loss function (cross entropy in the case of classification). The representation learned using an ODE solver is fed to the loss function which is optimized with respect to the  parameters $\theta$ 
$$\argmin_{\theta}  L(\mathbf{h}(T)) = \argmin_{\theta} L(\text{ODESolve}(\mathbf{h}(t_0), G, t_0, t_1, \theta))$$

The learning of the parameters requires back-propagating through the solver by computing the  gradients with respect to the loss, and this step is  computationally costly using naive back-propagation. \citet{NODE} proposed an adjoint sensitivity method to learn  the  parameters by running another  ODE solver backward in time. 

To optimize $L$ and the parameters $\theta$, we need to evaluate the  gradients with respect to $\mathbf{h}(t)$ (the state of our system at any time $t$), and $\theta$ the neural network parameters. The adjoint method describes a way to efficiently compute the derivative of the loss with respect to the state. 
In brief, we define the adjoint state as
\[
a(t) = -\partial L / \partial \mathbf{h}(t), 
\]
which describe the gradient of the loss with respect to some state $\mathbf{h}(t)$. It turns out that the  dynamics of the adjoint state can be described using another ODE.
\begin{equation}
\frac{da(t)}{dt} = - a(t)^{T} \frac{\partial G(\mathbf{h}(t), t, \theta)}{\partial \mathbf{h}}
\label{adjode}
\end{equation} 
The gradient  of the loss at the initial state $a(t_0)$ can be computed by running Eq.~\eqref{adjode} in the  backward direction with initial value as $a(T)$. We can compute the derivative of $G$ with respect to $\mathbf{h}$ easily by  computing the gradient through  back-propagation  in traditional neural networks. Now, the gradient of the loss with respect to parameters $dL/d\theta$ can be computed as
\begin{equation}
\label{eqn:loss_der}
\frac{dL}{d\theta} = - \int_{t_0}^T a(t)^\top \frac{\partial G(\mathbf{h}(t), t, \theta)}{\partial \theta } dt 
\end{equation}

The approach known as adjoint sensitivity has better memory cost, linear scalability  and low numerical  instabilities (refer ~\citep{NODE} for more details).

\subsection{NODE Adaptive Checkpoint Adjoint (ACA)}
\label{NODE_ACA}
The ``standard" NODE technique uses the adjoint method for learning the parameters by efficient back-propagation through the different numerical ODE solvers like Euler, Runge-Kutta, etc. But numerical errors prevail in the  computation of the gradient using the adjoint method ~\citep{NODE_ACA}, sometimes giving lower accuracies than expected. To mitigate this, Adaptive Checkpoint Adjoint (ACA) technique has been introduced,  which estimates more robust gradients for NODE.
We will refer to the NODE trained with ACA technique as NODE\_ACA in our paper. NODE\_ACA helps to achieve better accuracy by more accurate gradient calculation and lower computation time by removing the redundancy from the  computation graph.

NODE\_ACA ~\citep{NODE_ACA} saves forward pass and then applies this to backward pass, rather than backward trajectory being calculated independently of the forward pass as in the adjoint case.  The adjoint method does not maintain a history of the $h(t)$ computed in the forward pass but remembers the boundary conditions: $h(T)$ and $a(T)$. It then tries to solve $h(t)$ and $a(t)$ backwards in  time, i.e from $T$ to $0$ in order to compute the gradient of the loss function (\ref{eqn:loss_der}). However, due to numerical errors accumulated in the forward pass, $h(t)$ computed in the backward pass may not be accurate leading to the inaccurate computation of  the gradients as well as the  final solution.  On the other hand, in NODE\_ACA, discretization points $t_i$ and latent states $h_i = h(t_i)$ are recorded in the forward pass and reused in the  backward pass to reduce inaccuracies in the  gradient computation. This trajectory checkpoint strategy not only reduces numerical errors but also deletes shallow computation graphs. Both the constant and the adaptive stepsize solvers are supported by NODE\_ACA. Algorithm~\ref{algo:forward} provides the details of the adaptive step-size based numerical technique used in the forward pass.  Algorithm~\ref{algo:forward} summarizes the steps in the forward and backward passes of the  NODE\_ACA approach. NODE\_ACA stores the state values computed in the forward pass and uses them in the backward pass to make the gradient computation more accurate. 

\begin{algorithm*}[t]
\SetAlgoLined
  \SetKwInOut{Input}{Input}
  \SetKwInOut{Initialize}{Initialize}
    \Input{input data $h_0$, final time T, first stepsize $s_0$, error tolerance $etol$}
    \Initialize{h = $h_0$, s = $s_0$, error estimate \^{e} = $\infty$, t = 0}
    \While{t $<$ T}{
      \While{\^{e} $>$ etol}{
        s $\leftarrow$ s $\times$ decay\_factor(\^{e}) \\
        \^{e}, \^{h} = $\psi_s(t, h) \quad \quad $ // $\psi_s(t, h)$ compute the numerical solution at time $t+s$ 
      }
    t $\leftarrow$ $t + s$, h $\leftarrow$ \^{h}
    }
\caption{Numerical Integration algorithm with adaptive step-size used in the forward pass of Adjoint and adaptive checkpoint adjoint approaches.  }
\label{algo:forward}
\end{algorithm*}

\begin{algorithm*}[t]
\SetAlgoLined
  \SetKwInOut{Input}{Input}
  \SetKwInOut{Initialize}{Initialize}
  \SetKwInOut{ForwardPass}{ForwardPass}
  \SetKwInOut{BackwardPass}{BackwardPass}
  \Input{initial hidden state $h_0$, final time T, first stepsize $s_0$, error tolerance $etol$}
  \Initialize{h = $h_0$, s = $s_0$, error estimate \^{e} = $\infty$, t = 0}
\ForwardPass{ 
    \begin{enumerate}
    \item Perform numerical integration based on  Algorithm~\ref{algo:forward}. 
      \item Store discretization points ${t_0, ...t_{N_t}}$ and state  values ${h_0, h_1, ...h_{N_t}}$. 
      \item  Search for
            optimal stepsize by deleting local computation graphs 
    \end{enumerate}
}
\BackwardPass{Initialize $a(T)$, $dL/d\theta$ = 0}
 \For{$N_t$ to 1}{
  \begin{enumerate}
      \item  Compute $h\textsubscript{i+1}$ = $\psi_{s_i}(t_i, h_i)$ with stepsize $s_i = t\textsubscript{i+1} − t_i$
      \item Update $\lambda(t)$ and $dL/d\theta$ based on \eqref{adjode} and \eqref{eqn:loss_der}. 
      \item Delete local computation graphs
  \end{enumerate}
 }
\caption{Forward and Backward passes of the adaptive checkpoint adjoint (ACA) algorithm. Forward pass uses the  adaptive step size (Algorithm~\ref{algo:forward}) for numerical integration and state computation. The state values computed in the forward pass is reused in the backward pass. }
\label{algo:aca}
\end{algorithm*}

\section{Experimental Setup}
\label{EXPERIMENTAL SETUP}

\subsection{Network Architecture}
The network architecture used for the standard NODE training is as follows. We use a standard convolution block, consisting of two CLN (Convolution, Non-Linearity, Normalization) layers. Each convolution is done with a kernel of size 3$\times$3.

We also downsample the input, before passing it to the ODE network. Downsampling consists of applying 2-D convolutions, while reducing the number of channels. Once the input is down-sampled, it passes from the above ODE network, followed by a pooling layer. ODE maps the inputs to some desired latent space, which has the same number of dimensions as input. Similar to a classification task, as our final output has 5 dimensions (equal to the number of classes), we use a fully connected layer at the end. This FC (fully-connected) layer learns a linear mapping from the ODE output to the final output. 

For NODE\_ACA, we use the same architecture as that for standard NODE. Only difference being that, instead of using the adjoint method for back-propagation, the ACA technique is used.

For ResNet, we  use two NLC (Normalization, Non-Linearity, Convolution) layers for the architecture, instead of the CLN layers (as in the  standard NODE). This is also referred as Pre-Activation (as the ReLU operation is carried before convolution).  The convolution operation used here is again a 3$\times$3 convolution. Finally, the output from these layers is added to the original input (so that these layers only learn the residual). This constitutes one residual block. We take six such blocks, back to back, to form our ResNet. As mentioned above, down-sampling is applied before this network, followed by the pooling operation and FC layer at the end.

\subsection{Preprocessing}
Standard image processing is done on our image dataset similar to that in ~\citet{dai2018galaxy}, before it could be fed into the model. This is done so as to ensure that the images carry all the relevant information, needed  to accurately train the model. 

We mainly apply three image transformations. First, the image is resized  from $424 \times 424 \times 3$ pixels to $32 \times 32 \times 3$ pixels, using bilinear interpolation. This makes our training process faster as the number of dimensions in the  input image is largely reduced. The  transformation applied involves randomly flipping the image horizontally. The final transformation involves  image normalization, where all the three channels are normalized according to appropriate values. The data set is randomly split into training and testing set in the ratio 9:1 with  25911 images for training set and 2879 images for testing set as discussed in Section~\ref{DATASET}. We repeat this procedure 10 times and compute the  mean and variance over the evaluation metrics.  

\subsection{Implementation Details}

We use mini-batch gradient descent with a batch size of 256. 
The initial learning rate is
set to 0.1 and  then decreased by a factor of 10 to 30K and 60K
iterations. The
weight decay is set to 0.0001, dropout probability value to 0.8, and
the weights are initialized in the same way  as in \citet{DBLP:journals/corr/HeZRS15}. 

\subsection{Comparison of Computational Costs}
Here, we provide a brief  comparison of the  computational costs between ResNet, NODE, and NODE\_ACA.  To keep the comparison simple, as in \citet{NODE}, let  $L$ be the number of ResNet layers, and $\hat{L}$ be the number of forward-passes in NODE. The computational cost depends on $L$ and 
$\hat{L}$ for ResNet and NODE, respectively.
While $L$ is fixed and is  a  hyper-parameter, $\hat{L}$ is dependent on the  error-tolerance we set for NODE and NODE\_ACA. If the error-tolerance is high, $\hat{L}$ is comparable with $L$. If the error tolerance is low, then $\hat{L}$ is much higher than $L$. More details can be found in  Table 1 of \citet{NODE}.
\\
Similarly, while comparing $\hat{L}$ in the case of NODE and NODE\_ACA, $\hat{L}$ is roughly half in case of NODE\_ACA as compared to NODE. This is the reason why NODE\_ACA is roughly twice as faster than NODE. More details can be found in  Table 1 of \citet{NODE_ACA}.  For our analysis  for NODE\_ACA, it took about 12 hours to run (with about 11 hours for training and one hour for the testing)  a single network on a NVIDIA dgx server with P100 GPU. For NODE, it took about double the processing time and for RESNET it took about 90 minutes (80 minutes for training and 10 minutes for testing).

\section{RESULTS AND DISCUSSION}
\label{RESULTS AND DISCUSSION}

\subsection{Model Accuracy}

 Standard NODE model (trained with adjoint method) achieves an accuracy of 91-94\%, when trained with the Runge-Kutta method. The accuracy achieved by ResNet on similar architecture is between  89-94\%. With NODE\_ACA, we get accuracy between  91-95\%. Thus, we can say that NODE\_ACA achieves comparable accuracy  accuracy to ResNet, while having  one-third the  number of parameters. 
 For all the three networks, we get poor accuracy for  cigar-shaped images (class=2), due to the small number of images available for training. This is also consistent with the results in ~\citet{dai2018galaxy}.

 Table~\ref{tab:1} provides the confusion matrix for ResNet for different classes, while Table~\ref{tab:2} and Table~\ref{tab:3} provides the confusion matrix for standard NODE and NODE\_ACA, respectively, after averaging over the ten runs. The confusion matrix simply shows the contamination and completeness a particular category was classified into, among all the classes. It gives an idea, with which the other class, model confused a particular class the most. Data shown in the confusion matrices  were calculated, when the  output of the model led to the maximum correct predictions (or purity), when summed over all the classes. There is no other threshold which needs to be tuned for our problem.
  On the whole, the results of the completely round , the in-between, the edge-on and the spiral are extremely excellent, except for the cigar-shaped images (class=2). It happens due to the small number of  class=2  images for training.

\begin{table}[hbt!]
\label{confusion_matrix_table_resnet}
\centering
\begin{tabular}{c | c c c c c}
	&	0	& 1	& 2	& 3	& 4 \\
	\hline
0 & 757 & 57 & 0 & 0 & 29 \\
1 & 22 & 741 & 3 & 4 & 36 \\
2 & 0 & 12 & 11 & 27 & 5 \\
3 & 0 & 8 & 6 & 363 & 13 \\
4 & 7 & 12 & 1 & 13 & 748 
\end{tabular}
\caption{Confusion Matrix for   ResNet (averaged over all the 10 runs), where 0 : Completely round smooth, 1 : In-between smooth, 2 : Cigar-shaped smooth, 3 : Edge-on, and 4 : Spiral.}
\label{tab:1}
\end{table}

\begin{table}[hbt!]
\label{confusion_matrix_table_node}
\centering
\begin{tabular}{c| c c c c c}
	&	0	& 1	& 2	& 3	& 4 \\
	\hline
0 & 800 & 24 & 0 & 0 & 13 \\
1 & 36 & 700 & 0 & 8 & 22 \\
2 & 0 & 6 & 12 & 31 & 2 \\
3 & 0 & 5 & 7 & 365 & 8 \\
4 & 10 & 29 & 1 & 22 & 750
\end{tabular}
\caption{Confusion Matrix for NODE (averaged over all the 10 runs), where 0 : Completely round smooth, 1 : In-between smooth, 2 : Cigar-shaped smooth, 3 : Edge-on, and 4 : Spiral. }
\label{tab:2}
\end{table}

\begin{table}[hbt!]
\label{confusion_matrix_table_node_aca}
\centering
\begin{tabular}{c | c c c c c}
	&	0	& 1	& 2	& 3	& 4 \\
	\hline
0 & 786 & 36 & 0 & 0 & 12 \\
1 &  25 & 721 & 0 & 3 & 16 \\
2 &  0 & 4 & 23 & 20 & 1 \\
3 &  1 & 4 & 12 & 361 & 5 \\
 4 & 5 & 36 & 3 & 15 & 747 
\end{tabular}
\caption{Confusion Matrix for NODE\_ACA (averaged over all the 10 runs), where 0 : Completely round smooth, 1 : In-between smooth, 2 : Cigar-shaped smooth, 3 : Edge-on, and 4 : Spiral.}
\label{tab:3}
\end{table}

\subsection{Parameters Discussion}

{\tt ResNet} with 6 layers has 0.6 million parameters. Standard NODE on the other hand has total of 0.2 million parameters, for both Euler and Runge-Kutta ODE solving methods. NODE\_ACA network also has the same number of parameters as NODE. Thus, NODE and NODE\_ACA achieve similar overall accuracy with about one-third of the parameters as Resnet.

\subsection{Precision, Completeness (Recall), and F1}
We  compare the precision, Completeness (which is referred to as Recall in the Machine Learning Community), and F1 scores of standard NODE, ResNet, and NODE\_ACA. The Precision metric is the ratio of the total true positives  to  the total number of observations labelled positive. The Completeness tries to quantify what proportion of actual positives is correctly classified. While F1 is the harmonic mean of precision and completeness. More detailed definitions of these metrics can be found in ~\citet{Bethapudi}. These three metrics for Resnet, NODE, and NODE\_ACA can averaged over all the ten iterations can be found in  Table~\ref{table_accuracy_precision_recall_f1_all}. We can clearly see that the  performance of NODE and NODE\_ACA is comparable to that of  ResNet.

\begin{table*}[t]
  \centering
  \begin{tabularx}{\linewidth}{X| X X X| X X X| X X X| X X X}
  \hline
  Class &  \multicolumn{3}{c|}{Accuracy} & \multicolumn{3}{c|}{Precision}  & \multicolumn{3}{c|}{Completeness}  & \multicolumn{3}{c}{F1}	 \\
	\hline
	& ResNet & NODE & NODE \_ACA & ResNet & NODE & NODE \_ACA & ResNet & NODE & NODE \_ACA & ResNet & NODE & NODE \_ACA\\
	\hline
0 & $0.897\pm0.0061$ & $0.956\pm0.0008$ & $0.939\pm0.0007$ & $0.962\pm0.0029$ & $0.946\pm0.0007$ & $0.961\pm0.0012$ & $0.897\pm0.0061$ & $0.956\pm0.0008$ & $0.939\pm0.00071$ & $0.928\pm0.0029$ & $0.951\pm0.0003$ & $0.950\pm0.0006$ \\
1 & $0.917\pm0.0039$ & $0.914\pm0.0011$ & $0.941\pm0.0013$ & $0.892\pm0.0032$ & $0.917\pm0.0017$ & $0.896\pm0.0007$ & $0.918\pm0.0039$ & $0.914\pm0.0011$ & $0.942\pm0.0013$ & $0.905\pm0.0019$ & $0.916\pm0.0009$ & $0.918\pm0.0006$ \\
2 & $0.193\pm0.0294$ & $0.236\pm0.0050$ & $0.468\pm0.0092$ & $0.522\pm0.0491$ & $0.595\pm0.0202$ & $0.596\pm0.010$ & $0.193\pm0.0294$ & $0.236\pm0.0050$ & $0.468\pm0.0092$ & $0.281\pm0.0356$ & $0.338\pm0.0081$ & $0.524\pm0.0092$ \\
3 & $0.930\pm0.0064$ & $0.948\pm0.0007$ & $0.938\pm0.0011$ & $0.889\pm0.0063$ & $0.859\pm0.0020$ & $0.899\pm0.0013$ & $0.930\pm0.0064$ & $0.948\pm0.0007$ & $0.938\pm0.0011$ & $0.909\pm0.0041$ & $0.901\pm0.0013$ & $0.918\pm0.0006$ \\
4 & $0.957\pm0.0036$ & $0.9244\pm0.0011$ & $0.924\pm0.0008$ & $0.899\pm0.0037$ & $0.944\pm0.0007$ & $0.954\pm0.0008$ & $0.957\pm0.0036$ & $0.924\pm0.0011$ & $0.924\pm0.0008$ & $0.927\pm0.0024$ & $0.934\pm0.0006$ & $0.939\pm0.0006$ \\
\hline
\end{tabularx}
  \caption {\label{table_accuracy_precision_recall_f1_all} Accuracy, Precision, Completeness (Recall), F1 scores  for ResNet, NODE, NODE\_ACA along with error bars using all the 10 runs.}
  \label{tab:4}
\end{table*}

\subsection{ROC Curve}
\begin{figure}
\includegraphics[scale=0.55]{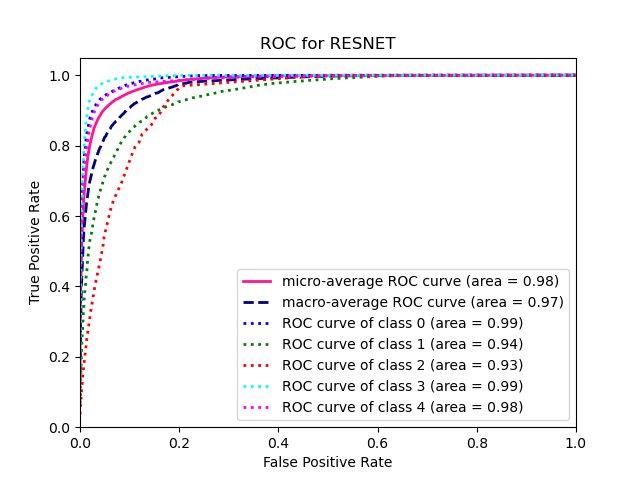}
\caption{ Average ROC curve for ResNet for the different classes. }
\label{roc_curve_resnet}
\label{fig:2}
\end{figure}

\begin{figure}
\includegraphics[scale=0.55]{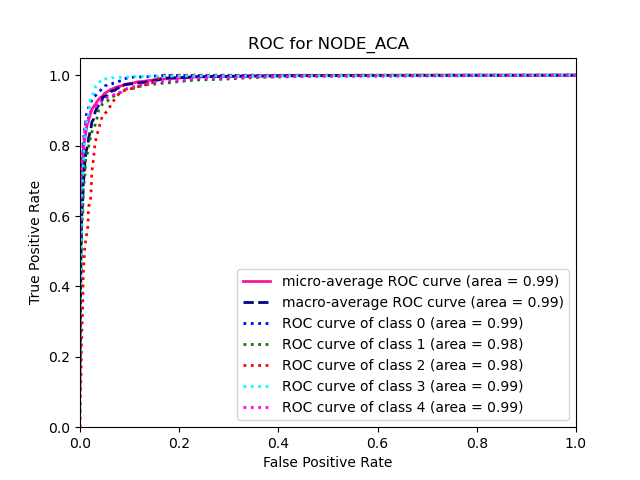}
\caption{Average ROC curve for standard NODE for the  different classes.}
\label{roc_curve_node}
\label{fig:3}
\end{figure}

\begin{figure}
\includegraphics[scale=0.55]{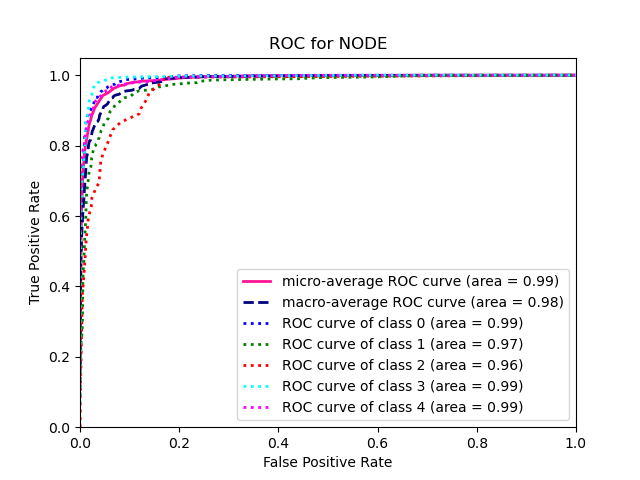}
\caption{Average ROC curve for NODE\_ACA for the different classes. }
\label{roc_curve_aca}
\label{fig:4}
\end{figure}

\begin{figure}[hbt!]
\includegraphics[scale=0.6]{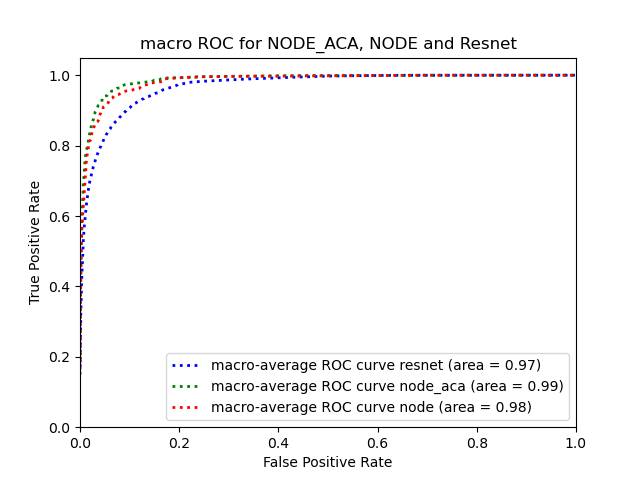}
\caption{Average Precision-Recall curves for ResNet, NODE, NODE\_ACA. }
\label{roc_curve_together}
\label{fig:5}
\end{figure}

ROC curve is an acronym for receiver operating characteristic curve. It plots the true positive against false positive rate, and shows how well a model is able to classify. Area under this curve is called AUC. The closer  AUC is to one, better is  the model in terms of classification. The ROC curves for each class, for  ResNet (Fig.~\ref{fig:2}), standard NODE (Fig.~\ref{fig:3}) and NODE\_ACA (Fig.~\ref{fig:4}) (after averaging over all the 10 runs) are shown. Micro and macro average for all the classes are also shown in the same figures. Micro-average is calculated by binarizing the output of each label, while macro-average is just the  unweighted average of each label. Thus, micro-average takes class-imbalance into account, giving more weightage to bigger classes while macro-average is forced to recognize each class correctly. These averages are well-versed in ML community~\citep{abdar2021uncertaintyfusenet}. As we can see, the  ROC curves for NODE for both the adjoint and the ACA techniques are very close to those of ResNet, for every image class. In Fig.~\ref{roc_curve_together}, we plot the average curve (averaged over all the  classes), for all the  three aforementioned techniques. As we can see, the performance of NODE and NODE\_ACA  is comparable to  ResNet.


\section{CONCLUSIONS} \label{CONCLUSIONS}
In this paper, we have used NODE with adjoint training and NODE\_ACA technique  for the task of galaxy morphology classification, and also compared its performance with ResNet. The dataset used for this purpose is a subset of Galaxy Zoo 2 dataset. We find that  the number of parameters in NODE (standard and ACA) is about one-third compared to its counterpart ResNet, and it achieves this without compromising the performance. Accuracy achieved by NODE (with adjoint training) is the same as ResNet, while NODE (with ACA technique) achieves an accuracy of 91-95\%, \rthis{where the ground truth is determined by the Galaxy Zoo 2  classifications.}
Also, with both the NODE techniques, we can easily trade-off accuracy for speed, which is not possible for ResNet.


We also compare all the three models using other metrics such as precision, Completeness, F1, and AUC. Through our results, we conclude that, the performance of  NODE and NODE\_ACA   is comparable to  ResNet, for all these metrics, while providing all the advantages guaranteed by ResNet. We also illustrate this, by plotting the average performance of standard NODE, ResNet and NODE\_ACA, on one graph, as shown in Fig~\ref{roc_curve_together}.

From our experiments, we therefore conclude that NODE has several advantages over ResNet and  can easily supersede ResNet. With large scale astronomical surveys coming up, and more and more data being generated from these surveys, there is  a pressing need to replace such classification tasks with robust deep learning  models. These emerging deep learning  models can not only help  speed-up the process of training and  classification, but also provide better insights, by breaking down the process in series of small steps. Thus, researchers have better control over the process, and can easily trade-off one parameter (like accuracy) with another (like speed). Our methodology would prove to be beneficial for upcoming large scale astronomical surveys such as Vera Rubin LSST, Euclid, WFIRST etc.

All our codes used for the analysis in this work are publicly available at \href{https://github.com/rg321/torch_ACA}{github.com/rg321/torch\_ACA}. We also provide some rudimentary guidance on how to use both NODE and NODE\_ACA for the supervised classification problem in the appendices.

\section{ACKNOWLEDGEMENTS}
\label{Ack}

We would like to thank the galaxy challenge, Galaxy Zoo,
SDSS and Kaggle platform for sharing their  data. RG is supported by funding from  DST-ICPS (T-641). \rthis{We are grateful to the anonymous referee for useful feedback on our manuscript.}



\nocite{*}
\bibliographystyle{model2-names}
\bibliography{example_paper}
\newpage

\appendix
\section{Instructions for using NODE\_ACA within PyTorch}
We provide some bare-bones guidelines on how to access and use  NODE\_ACA for any  classification problem. NODE\_ACA is available in both TensorFlow and PyTorch.
For this work, we have used PyTorch and hence provide some  a rudimentary guide on the usage of  NODE\_ACA  in  PyTorch.
Our full analysis has also been provided in a github link at \href{https://github.com/rg321/torch_ACA\_gz}{github.com/rg321/torch\_ACA\_gz}. 

\citet{NODE_ACA} provide a PyTorch package at \url{https://github.com/juntang-zhuang/torch_ACA}, which can be easily plugged into the existing models, with support for multi-GPU training and higher-order derivative.
A simple way to plug NODE\_ACA into your existing code is as follows.
\begin{verbatim}
from torch_ACA import odesolve_adjoint as odesolve
out = odesolve(odefunc, x, options)
\end{verbatim}

One then needs to write a   custom data-loader in order to load the data into the model. For example, to load the Galaxy Zoo data images into the model for training and testing purposes, a custom data-loader $get\_gz\_loaders$ in file $data\_loader.py$ is written and used. This loader is written in PyTorch's standard \href{https://pytorch.org/docs/stable/data.html#torch.utils.data.DataLoader}{DataLoader} style. It fetches  the images using PyTorch's \href{https://pytorch.org/vision/stable/datasets.html#torchvision.datasets.ImageFolder}{ImageFolder} function, apply necessary transformations, splits them into training and testing parts and finally returns $train$ and $test$ $DataLoader$, which are standard PyTorch objects used for data loading.

Once the data-loaders are in place, rest of the flow is the  same as for any other dataset like \href{https://www.cs.toronto.edu/~kriz/learning-features-2009-TR.pdf}{CIFAR10}, ImageNet \citep{russakovsky2015imagenet} etc. The only thing that needs to be done now is tuning the hyper-parameters in order to get the best accuracy or whatever desired. For example, to run it on galaxy-zoo datset, run the following command -:
\begin{verbatim}
python train.py --num_epochs 15 --dataset galaxyzoo 
--batch_size 64 --test_batch_size 32
\end{verbatim}

An example usecase of NODE\_ACA in the TensorFlow library can be found in \url{https://github.com/titu1994/tfdiffeq}

\section{Instructions for using NODE within PyTorch}
After creating the DataLoader as described in above section, use the following command (on the  Linux prompt)  to use NODE architecture on your data (for example, dataset used is MNIST here) -:
\begin{verbatim}
 python train.py  --data galaxyzoo   
 --optimizer sgd   --lr 0.1 --solver  runge_kutta   --use_ode
\end{verbatim}
\end{document}